\documentstyle[12pt]{article}
\input{psfig}
\newlength{\extraspace}
\newlength{\extraspaces}
\newcommand{\beq}{\begin{eqnarray}
\addtolength{\abovedisplayskip}{\extraspaces}
\addtolength{\belowdisplayskip}{\extraspaces}
\addtolength{\abovedisplayshortskip}{\extraspace}
\addtolength{\belowdisplayshortskip}{\extraspace}}
\newcommand{\eeq}{\end{eqnarray}}
\newcommand{\bea}{\begin{eqnarray}
\addtolength{\abovedisplayskip}{\extraspaces}
\addtolength{\belowdisplayskip}{\extraspaces}
\addtolength{\abovedisplayshortskip}{\extraspace}
\addtolength{\belowdisplayshortskip}{\extraspace}}
\newcommand{\eea}{\end{eqnarray}}

\newcommand{\goto}{\rightarrow}
\newcommand{\be}{\beta}

\newcommand{\al}{\alpha}

\newcommand{\om}{\omega}
\newcommand{\xvec}{{\bf x}}

\newcommand{\kvec}{{\bf k}}
\renewcommand\slash{\!\!\!/}

\begin{document}
\addtolength{\baselineskip}{.8mm}
\thispagestyle{empty}

\begin{flushright}
OU-TP-98-61P\\
{ hep-th/9808068}\\
{ July 1998}
\end{flushright}
\vspace{0.5cm}
\begin{center}

{\large\sc{VACUUM INSTABILITY IN TOPOLOGICALLY MASSIVE GAUGE 
THEORY}  }\\[15mm].

{Alex
Lewis\footnote{e-mail:
a.lewis1@physics.oxford.ac.uk} \\}
\vspace{0.5cm}
{\it   Department of Physics, Theoretical Physics \\
  University of Oxford, 
1 Keble Road, Oxford, OX1 3NP \\ United Kingdom}\\
\vspace{0.5cm}
{\sc Abstract}
\end{center}
\noindent
{ We find the critical charge for a topologically massive gauge theory for any
gauge group, generalising our earlier result for SU(2). 
The relation between critical charges in TMGT, singular
vectors in the WZNW model and logarithmic CFT is investigated. }
\vfill

\newpage
\pagestyle{plain}
\setcounter{page}{1}

\section{Introduction}

It is  well known that a topological Chern-Simons theory 
 on a
3-dimensional manifold with a boundary induces a WZNW model 
 on the boundary \cite{witten} which is a basic ``building block''
 for  all known unitary rational conformal field theories (CFT).
  Combining several
Chern-Simons fields and/or factorising over some discrete symmetries
 one can  give a three-dimensional construction for 
 all known unitary rational CFT \cite{MS}, for example,
minimal models \cite{bpz} through a GKO coset construction \cite{gko}.
In unitary WZNW and minimal models, primary fields only 
exist for a restricted number of
representations.  For example for  the   $SU(2)$ model,
  only the representations
with $j = 0,\frac12,\dots,\frac{k}{2}$ are allowed, while in the
minimal model the allowed  primary fields are those which satisfy the
above condition for each of the three $SU(2)$ factors in the GKO
construction. In \cite{us} we showed that the truncation of the
spectrum in the $SU(2)$ case was associated with a critical charge in
the corresponding topologically massive gauge theory (TMGT). This let
us see that critical and super-critical charges would cause vacuum
instability and be excluded from the spectrum of the three-dimensional
theory as well. We also found that, surprisingly,
 the quantization of a theory with a
critical charge, which was worked out in \cite{ss}, leads to a Jordan
block structure for the Hamiltonian which is similar to what is now
known from logarithmic conformal field theory. 
The  purpose of
this paper is to present these results in detail and extend them
 to any compact group.

The topological
  Chern-Simons theory  is the low-energy limit of a topologically massive 
 gauge theory \cite{ziegel}-\cite{tmgt}  with  the action:
\bea
S_{\tiny TMGT} &=& -\frac{1}{2e^2} \int_{\cal
M}d^3x~\mbox{tr}~F_{\mu\nu}F^{\mu\nu}+ kS_{\tiny CS}
\label{action}\\
S_{\tiny CS} &=& \int_{\cal M} {1\over4\pi}~\mbox{tr}\left(A\wedge dA +
{2\over3}A\wedge A\wedge A\right)=\int_{\cal
M}d^3x~\frac{1}{4\pi}\epsilon^{\mu\nu\lambda}~\mbox{tr}\left(A_\mu
\partial_\nu
A_\lambda+\frac{2}{3}A_\mu A_\nu A_\lambda\right)
\nonumber\eea
Where $A_\mu=A_\mu^at^a$, and $t^a$ are the generators of the gauge
group $G$. This is the action for gauge bosons with topological mass
$M=ke^2/4\pi$. If the three dimensional manifold $\cal M$ has a
boundary, the TMGT induces a deformed conformal field theory on the
boundary. In the nonabelian case the CFT  will be
deformed with a deformation parameter proportional to $1/e^2$, and so
 the WZNW model can be induced from a TMGT with a  boson  mass 
of the  order
 of the UV cut-off, so that the $F^2$ term regularizes the Chern-Simons
action and we recover the WZNW model by letting the 
mass $M\goto \infty$. If the TMGT is defined on a manifold with a
boundary, it will induce a chiral WZNW model on the boundary. By
taking a cylinder, with one chiral half of the WZNW model on each
boundary, one can obtain the full WZNW model, defined by
\beq
S_{\tiny WZNW}(k;g)={k\over8\pi}\int d^2z~\mbox{Tr}g^{-1}\partial^\mu
gg^{-1}\partial_\mu g~+~{ik\over12\pi}\int
d^3z~\mbox{Tr}g^{-1}dg\wedge
g^{-1}dg\wedge g^{-1}dg
\label{3dwzw}\eeq
A primary field of the WZNW model,
$\Phi(z,\bar{z})=V_L(z)V_R(\bar{z})$, can be induced on the boundary by
a path-ordered Wilson line
\beq
W_R[C] = \mbox{tr}_R \exp{\left(i\int_C A\right)}
\eeq
where $C$ is a contour with one end point on each boundary. The Wilson
line coincides with The holomorphic part of the primary field $V_L$
on one boundary and with the anti- holomorphic part $V_R$ on the
other, and it describes the transport of a particle in the
representation $R$ from one boundary to the other. It is therefore
clear that the same representations of the gauge group have to occur
in the bulk as on the boundary.

In the same way, a supersymmetric WZNW model can be induced on the
boundary of a three dimensional manifold by a supersymmetric CS
theory, which is the low energy limit of a supersymmetric TMGT,
defined by the action \cite{ziegel,st}
\beq{\begin{array}{lll}
{\cal S}&=&S_{\tiny TMGT}+\frac{1}{2e^2}\int
d^3x~\bar\chi^ai\gamma^\mu\left(\partial_\mu\chi^a+if^{abc}A_\mu^b\chi^c
\right)-\frac{k}{8\pi}\int d^3x~\bar\chi^a\chi^a\\& &~~~+\frac{k}{8\pi}\int
d^3x~i\bar\lambda^a\left(\gamma^\mu\partial_\mu\chi^a-\frac{1}{3}f^{abc}
\gamma^\mu\partial_\mu\lambda^bL^c-\frac{2}{3}f^{abc}\epsilon^{\mu\nu\rho}
\gamma_\nu\partial_\mu\lambda^bA_\rho^c\right)\end{array}}
\label{susytmgt}\eeq
where $\chi^a$ are Majorana fermion fields in the adjoint representation of
$G$, and $L^a$ are auxiliary scalar fields with $\lambda^a$ their Majorana
spinor superpartner fields. In this case we have both vector bosons
and Majorana fermions with the topological mass $M=ke^2/4\pi$ in the
bulk, and on the boundary we get the super-WZNW model with the action
\cite{vkpr}
\beq
S_{\tiny SWZNW}=S_{\tiny WZNW}(k;g) +  \frac{ik}{8\pi}\int d^2z~ 
\mbox{tr} \left[
\bar\psi^\dagger \left[\partial\slash +
\gamma_5\partial\slash gg^\dagger \right] \psi 
\right]
\eeq
The fermions in the SWZNW action can be decoupled completely by the
transformation
\beq
\chi = \frac{1}{2}\left[ (1+\gamma_5) g^\dagger\psi + (1-\gamma_5)\psi
g^\dagger \right]
\eeq
but this transformation is anomalous, and leads to the following
action
\beq
S_{\tiny SWZNW}=S_{\tiny WZNW}(k;g-C_2) +  \frac{ik}{8\pi}\int d^2z~
\mbox{tr} \left[ \bar\chi\partial\slash\chi\right]
\eeq
where $C_2$ is the Casimir of the adjoint representation of the group
$G$
\beq
C_2\delta^{ab} = f^{acd}f^{bcd}
\eeq
The bosonic part of the super-WZNW model at level $k$ is therefore the
ordinary WZNW at level $k-C_2$, and the spectrum  in
the supersymmetric model is the same as for the WZNW at  
level $k-C_2$ and  the free fermions. For example, the conformal
dimension of a primary field in a representation ${\cal R}$ of $G$,
with Casimir $C_R$, is $\Delta = C_R/(k+C_2)$ in the WZNW model and
$\Delta = C_R/k$ in the SWZNW model. In the three-dimensional
theories, $1/k$ is the expansion parameter for perturbation theory, so
the ordinary TMGT and the bosonic part of the SUSY TMGT are the same
at tree level, with the shift $k \goto k-C_2$ appearing as a result of
higher-order corrections \cite{acks}. In this paper we will only be
working at the leading order in $1/k$ and so we will not be able to
see the
difference between the ordinary and supersymmetric TMGTs.

\section{Truncation of the spectrum in WZNW model and in TMGT}

We will begin by reviewing how the truncation of the spectrum of
unitary representations occurs in the WZNW model, and then see how the
same spectrum can arise by a completely different mechanism in TMGT.
The WZNW has a $G_L \times G_R$
symmetry, generated by the currents
\bea
J(z)=J^a(z)t^a &=& -\frac{k}{2} \partial g g^{-1} \nonumber \\
\bar{J}(\bar{z})=\bar{J}^a(\bar{z}) &=&
-\frac{k}{2}g^{-1}\bar{\partial} g
\eea
The modes $J^a_n$ generate the affine algebra with the commutation
relations
\beq
[J^a_n,J^b_m] = \frac{nk}{2}\delta^{ab}\delta_{n+m,0}
+f^{abc}J^c_{n+m}
\label{KM}\eeq
We use the following basis for the generators of $G$. We have the
generators of the Cartan subalgebra $h^i$, $i=1,\dots,r$, $r$ is the
rank of $G$, and for each root $\al$ we 
have a step operator $e^\al$,
with the commutation relations 
\bea
\left[h^i,h^j\right] &=& 0 \nonumber \\
\left[h^i,e^{\al}\right] &=& \al^i e^{\al} \nonumber \\
\left[e^\al,e^\be\right] &=& 
\epsilon(\al,\be)e^{\al+\be}~~\mbox{if $\al+\be$ is
a root} \nonumber \\
&=&  2\al\cdot h/\al^2~~\mbox{if $\al =- \be$} \nonumber \\
&=& 0~~ \mbox{otherwise}
\label{gbasis}\eea
where $\epsilon(\al,\be)$ is antisymmetric in $\al$ and $\be$, and
$\al\cdot h = \sum_i \al^i h^i$. 
The current  algebra of the WZNW model
then has generators $H^i_n$, $E^\al_n$, and the commutation relations
(\ref{KM}) become
\bea
\left[H^i_m,H^j_n\right] &=& \frac{mk}{2}
\delta^{ij}\delta_{n,-m} \nonumber \\
\left[H^i_m,E^{\al}_n\right] &=& \al^i E^{\al}_{m+n} \nonumber \\
\left[E^\al_m,E^\be_n\right] &=& 
\epsilon(\al,\be)E^{\al+\be}_{m+n}~~\mbox{if $\al+\be$ is
a root} \nonumber \\
&=&  \left\{2\al\cdot H_{m+n} +km\delta_{m,-n}\right\}/
\al^2~~\mbox{if $\al =- \be$} \nonumber \\
&=& 0~~\mbox{otherwise}
\label{KMcoms}\eea
We consider a highest weight primary state $|\mu\rangle$, defined by
\bea
H^i_n|\mu\rangle = E^\al_n|\mu\rangle &=& 0,~~~\mbox{for $n\geq 1$} 
 \nonumber \\
H^i_0|\mu\rangle &=& \mu^i|\mu\rangle \nonumber \\
E^\al_0|\mu\rangle &=& 0,~~~ \mbox{for $\al$ a positive root}
\label{hwstate}\eea
Unitary representations only exist for weights $\mu$ which satisfy
$\psi\cdot\mu\leq k/2$ with the highest root $\psi$ normalized so that
$\psi^2=1$.  
This can easily be seen by
considering the norm of a descendant, using eqs.
(\ref{KMcoms}) and (\ref{hwstate}) and $E_n^{\al\dagger}=E_{-n}^{-\al}$:
\beq
|E^\psi_{-1}|\mu\rangle |^2 = \langle\mu
|E^{-\psi}_1E^\psi_{-1}|\mu\rangle =
\frac{1}{\psi^2}\left(k-2\psi\cdot\mu\right)\langle\mu|\mu\rangle
\eeq 
This state  therefore has negative norm when $\psi\cdot\mu >k/2$, and it is a
null vector for a highest weight $\mu_0$ such that  $\psi\cdot\mu_0
=k/2$. 
This null vector has all the
properties (eq. (\ref{hwstate})) 
of a primary state with highest weight $\mu_0+\psi$, and so we can
write $||\mu_0+\psi\rangle |^2 =0$. There are also singular vectors at
higher levels for the other unitary representations. In a unitary
model, any correlation function containing a singular vector must be
zero, and this leads to all non-unitary representations decoupling
from the spectrum \cite{gw}.

It is clear that, if the CFT is to be described by a three-dimensional
Chern-Simons theory, this truncation of the spectrum also has to occur
in the CS theory, and there must be states with zero norm in the CS
theory as well. This must also be true
 at least in the low-energy limit in the TMGT
theory, but in fact, we will see that it actually 
occurs in the full TMGT theory.
To induce a primary field of the WZNW model on the boundary in a
representation ${\cal R}$, we have to add to the CS model in the bulk 
a Wilson line carrying a
source in the same representation, and so we expect that adding a
source in a representation with $\mu\cdot\psi > k/2$ would for some
reason make the three-dimensional theory non-unitary.
In
\cite{us} we suggested that the physical reason for this is that
sources with higher charge are critical or super-critical - ie. that
they cause vacuum instability because the bound states which they can
form
with 
charged particles  have zero or imaginary energy. This is just the
sort of behaviour that is familiar in the solution of 
the Dirac equation for an
electron in the field of a point charge $Ze$  in which case  the energy
of the ground state is 
$E(Z) = m_e \sqrt{(1-\al^2Z^2)}$, and
so there is a critical charge $Z_c=1/\al \approx 137$, with $E(Z_c)=0$
and imaginary energy for $Z>Z_c$. In a nonabelian TMGT, the charged
particles which can have bound states are just the massive
gauge bosons themselves.
We can therefore see that the dynamics of the TMGT are crucial for
understanding even the low energy properties of the theory -- even when
the gauge bosons are infinitely massive, any states with exactly zero
energy obviously cannot be ignored, and super-critical charges will
never be formed because they would always be screened by pairs of
charged
particles which could condense from the vacuum with no energy cost. 
We will also see in the next section that the presence of a critical
charge leads to a state with zero norm, just as in the WZNW model.

In this section we present the computation of the critical charge
in a TMGT \cite{us}, showing how the result is generalized to
arbitrary groups. We divide the gauge bosons into uncharged
``photons'' and charged bosons, by
\beq
A_\mu = \sum_i A_\mu^ih^i +\sum_\al A_\mu^\al e^\al
\eeq
where the fields $A^\mu_i$ describe the photons and $A^\mu_\al$ the
charged particles. The charged bosons can be divided into positively
and negatively charged for positive and negative roots -- for a 
positive root $\al$, $\al^i \geq 0$ for all $i$. Since $-\al$ is
always a root if $\al$ is a root, the charged bosons
therefore come in particle-antiparticle pairs with positive and
negative charges. A classical external source in a highest weight
state will then act as a source of photons, creating a potential for
the charged particles. We will 
therefore treat the photons classically and
try to solve the equations of motion for charged particles in the
classical background created by the source.
The  equations of motion for the photons in the presence of a static
source with weight $\mu$ are
\beq
\frac{1}{2}\epsilon^{\mu\nu\lambda}F^i_{\nu\lambda}
+ \frac{1}{M} \partial_\nu F^{i,\nu\mu} +
\frac{2\pi \mu^i}{k}\eta^{0\mu}\delta^2(x) = 0
\label{sources}\eeq
This leads to the following solution for the electric and magnetic
fields, $E^i_j = F^i_{j0}$ and $B^i=F^i_{12}$: 
\beq
B^i(r) = -\frac{\mu^iM^2}{k}K_0(Mr),~~~~~E^i_\theta(0) = 0,~~~~~
E^i_r(r)= -\frac{\mu^iM^2}{k}K_1(Mr)
\label{bground}\eeq
where $K_0$ and $K_1$ are modified Bessel functions.
By integrating eq. (\ref{sources}) we also find that at large
distances ($r \gg 1/M$)
\beq
\int B^i d^2x = -\frac{2\pi\mu^i}{k}
\eeq 
so that, by Stokes' theorem, $A^i_\theta \sim -\mu^i/kr$.
The fields $A^i_\mu$ 
can then conveniently be written in the gauge
$\partial_\mu A^\mu = 0$ as
\beq
A^i_0 = -\frac{B^i}{M},~~~~~~A^i_r = 0,~~~~~A^i_\theta = \frac{E^i_r}{M}
 - \frac{\mu^i}{kr}
\label{Ai}\eeq
We could now go on to write down the linearized equations of motion
for the charged bosons in this background field, and try to solve them
to find the energy of the ground state and hence the critical charge. 
In fact it is much simpler to
analyse the equations for fermions in the same background. For this
reason, it is useful to consider the supersymmetric TMGT, 
 The background gauge field in the presence of a classical source is
the same in the SUSY case as in the ordinary TMGT, and the spectrum of
the bosons and fermions in the SUSY theory is of course the same, so
to find the critical charge we can concentrate on the equation of
motion for the fermions, which is just the Dirac equation
 $(i\gamma^\mu D_\mu
-M)\chi=0$. 
As with the bosons, we can divide the fermions into charged and
neutral particles by
\beq
\chi = \sum_i \chi^ih^i +\sum_\al \chi^\al e^\al
\eeq
where the charge is positive if $\al^i \geq 0$, so that $\al\cdot\psi
>0$. 
We use a
 Majorana representation for the gamma
matrices in $2+1$ dimensions 
$\gamma_0=\sigma_2,\gamma_1=i\sigma_1,\gamma_2=i\sigma_3$, where
$\sigma_i$ are the Pauli matrices.
The Dirac equation for a charged fermion is then
\beq
\left( \begin{array}{cc} -D_2 -M & -D_0-D_1 \\
D_0-D_1 & D_2-M \end{array} \right) \left(\begin{array}{c}  
\chi^\al_1 \\ \chi^\al_2 \end{array} \right) =0
\label{Dchi}\eeq
where from the commutation relations (\ref{gbasis}) we can see that
$D_\mu \chi^\al = (\partial_\mu + i\al^i A^i_\mu) \chi^\al$.
We would expect a
bound state for negatively charged fermions for a source with
$\mu\cdot\psi >0$, but in some circumstances in a TMGT there can be
bound states between like charges and repulsion between unlike charges
\cite{dekss}. This is because a charge is a source of both electric
and magnetic flux and the magnetic force is in the opposite direction
to the usual electric force. We need to check that this is not the
case in the present situation, as if there is no attraction between
opposite charges there can obviously be no screening of critical
charges. An easy way to check this is to take the non-relativistic
limit of eq. (\ref{Dchi}). We define  
 $\chi^\al_\pm = \chi^\al_1 \pm \chi^\al_2$, and in the
non-relativistic limit  the total energy $E=M+\epsilon$ with 
$|\epsilon| \ll M$ and also $|A^i_0| \ll M$. In this
limit the Dirac equation becomes the Pauli equation, which in this
case is
\bea
\chi_\pm &=& \chi_\pm e^{i(M+\epsilon)x_0} \nonumber \\
\left(\epsilon + \al^i A^i_0\right) \chi_+ &=& 
\frac {-i}{2M}\left( (D_i)^2 -i\al^iB^i\right) \chi_+
\label{pauli}\eea
The first term on the right of eq. (\ref{pauli}) is the usual kinetic
energy term, except that the centrifugal energy is modified by the
Aharonov-Bohm effect, and the second term is the magnetic
interaction. From eq. (\ref{Ai}) we can see that the strength of the 
magnetic energy is just half that of the electric interaction, and so
we still have attraction between opposite charges.

Since we are really interested in the critical charge for which
$\epsilon =-M$, we cannot use the non-relativistic approximation any
further and so we return to eq. (\ref{Dchi}). the attraction will be
strongest for the negatively charged fermion $\chi^{-\psi}$, 
and so we need only consider
that.
If we define
$q=\psi\cdot\mu/k$
and make the gauge transformation
\beq
\chi \goto \chi' = \chi \exp\left[-\frac{A_0(r)}{M} - q \log(r) 
\right]
\eeq
and define, for a solution with angular momentum $m$ and energy $E$
\bea
x_\pm  &=& x_1 \pm i x_2 \nonumber \\
f(r) &=& -i x_+^{-m} \chi^{-\psi}_+ e^{iE x_0}\nonumber \\
g(r) &=& r x_+^{-(m+1)} \chi^{-\psi}_- e^{iE x_0}
\eea
we obtain the following equations (with $x=Mr$, $\om=E/M$)
\bea
&&\left(1 - \om - q K_0(x) \right) g(x) + f'(x) = 0 \label{fgeqns} \\
&&\left(1 + \om + q K_0(x) \right) f(x) + 
\left( (2q+2m+1)x^{-1} -2qK_1(x) \right)g(x) +g'(x) = 0
\nonumber\eea
These are exactly the equations which we solved numerically in
\cite{us}, and we can now see that with the above definition of $q$ 
they apply 
for any simple group. 
We found the expected behaviour, with a critical charge
very close to $q=0.5$ (actually 0.498), because of which we conjecture
that the exact solution will have the critical charge at exactly $q=0.5$
This appears to indicate that we do indeed have the same truncation of
the spectrum in TMGT as in the WZNW model, with only representations
with $\psi\cdot\mu \leq k/2$ allowed. However, this calculation is
really only accurate to the lowest order in $1/k$, so we should say
the critical charge is at 
\beq
\psi\cdot\mu = \frac{k}{2}\left\{1+O(k^{-1})\right\}
\eeq
Of course, in the supersymmetric models we have to shift $k \goto k-C_2$
compared to the bosonic models \cite{vkpr,fuchs,acks}, but clearly
this cannot be seen at this order. Also, we would not expect a
critical charge at $\psi\cdot\mu =k/2$ as this representation is 
included in the physical spectrum of
 the WZNW model, and we would expect a critical charge to be
screened a particle-antiparticle pair which could condense from the
vacuum.
 We might expect that the true critical
charge is $\psi\cdot\mu =(k+1)/2$ which would exclude all the
non-unitary representations since $\psi\cdot\mu$ is always an integer
or half-integer, or that it is $\psi\cdot\mu =(k+2)/2$, which is the
null state in the WZNW model (with $k$ replaced by $k-C_2$ in the
supersymmetric 
case). 
To decide between these possibilities, we need to know how to quantize
a theory with a critical charge -- it turns out that the existence of a
critical charge leads to a state having zero norm, so 
that the latter is probably
correct, but it is not yet clear why states with $\psi\cdot\mu
=(k+1)/2$ should also be screened.

\section{Critical States}

The quantization of
fields with zero  energy bound states  has   been  worked
out in \cite{ss}. Although we have 
found a bound state for fermions, in the supersymmetric theory the
spectrum for the gauge bosons must be the same, and so there will be a
critical bound state of bosons at the same critical charge, and we can
also expect the same features in the non-supersymmetric case, with the
shift of $k \goto k+C_2$. Since the bosons exist in all models we
expect that the features in which we are interested will all appear in
the quantization of a theory with a critical state for charged bosons,
and we concentrate on that case.
The treatment in \cite{ss} 
 was for scalar bosons, but as vector bosons in
$2+1$ dimensions only have one degree of freedom we can follow
it exactly.
We have the canonical
commutation relations
\bea
\left[ A({\bf x}),\Pi({\bf y})\right] = 
i\delta({\bf x}-{\bf y}) \nonumber \\
\left[ A({\bf x}),A({\bf y})\right]=0,~~\mbox{etc.}
\label{3dcanonical}\eea
We expand the field $A$ and the conjugate momentum $\Pi$ in
terms of wave functions $\Phi^\kvec(\xvec)$ for continuum states with
$\om^2=\kvec^2+1$ and $\Phi^i(\xvec)$ for bound states with $\om_i^2 < 1$,
normalized so that
\bea
\int {\bar\Phi}^i(\xvec)\Phi^i(\xvec) d^2x=1 \nonumber \\
\int {\bar\Phi}^\kvec(\xvec) \Phi^{\kvec'}(\xvec) d^3x
=\delta(\kvec - \kvec')
\eea
The field operators at $t=0$ can be expanded as
\bea
A(\xvec) &=& \sum_i q_i\Phi^i(\xvec) + 
\int q(\kvec)\Phi^\kvec(\xvec)d^3k \nonumber \\
\Pi(\xvec) &=& \sum_i p_i{\bar\Phi}^i(\xvec) + 
\int p(\kvec){\bar\Phi}^\kvec(\xvec)d^3k
\eea
The commutation relations (\ref{3dcanonical}) then become 
\beq
[q_i,p_i]=i,~~~~~[q(\kvec),p(\kvec')]=i\delta(\kvec-\kvec'),~~~~~
[q_i,q_j]=0,\mbox{etc.}
\eeq
For bound states with $0<\om_i^2<1$, we introduce the mode operators
$a_i$ and $b_i$ according to the standard rule
\bea
q_i &=& \left(\frac{1}{2\om_i}\right)^{1/2}(b_i^\dagger + a_i)
\nonumber \\
p_i &=& -i\left(\frac{\om_i}{2}\right)^{1/2}(b_i - a_i^\dagger)
\label{modes}\eea
In terms of these operators, when there are no bound states with zero
or imaginary energy the Hamiltonian is
\beq
H=  \sum_i  (p_i^\dagger p_i +  \om_i^2q_i^\dagger q_i) +H_C
 =  \sum_i  (a_i^\dagger a_i + b_i^\dagger b_i) +H_C
\eeq
where $H_C$ is the part of the Hamiltonian that comes from the
continuous states. The states $|a_i\rangle \equiv a_i^\dagger
|0\rangle$
 and $|b_i\rangle \equiv b_i^\dagger
|0\rangle$ satisfy
\bea
H|a_i\rangle & =& \om_i |a_i\rangle, ~~~~~~~~~~~~~~\langle a_i | a_i
\rangle = 1  \nonumber \\
H|b_i\rangle & = &\om_i |b_i\rangle, ~~~~~~~~~~~~~~\langle b_i | b_i
 \rangle  = 1
\eea
So the Hamiltonian is of course diagonalizable and all states have
positive norm.

Now we turn to the situation when there is a critical bound state,
 with $\om_0=0$. The contribution of the critical state 
to the Hamiltonian is just
\beq
H_0 = p_0^\dagger p_0
\eeq
In the low energy limit we can take this to  be the full Hamiltonian. 
Now we
need to be careful about exactly how we quantize the theory. As we
will see, there are actually two choices we can make, one of which
leads to a unitary CFT on the boundary and the other to a logarithmic
CFT.
Instead of eq. (\ref{modes}), 
in this case we introduce the $p_0$ and $q_0$  as mode
operators
\beq
c=p_0,~~~~~~~d=-iq_0^\dagger
\eeq
with the commutation relations 
\beq
[c,c^\dagger]=[d,d^\dagger]=0,~~~~~~~~~[d^\dagger,c]=-1
\eeq
The Hamiltonian is $H = c^\dagger c$.  Starting from a naive vacuum
state $|0\rangle$, which of course is really a state containing a
critical charge in the full TMGT,  we can make  two more zero energy 
states,
$|c\rangle = c^\dagger |0\rangle$ and $|d\rangle =(c^\dagger +
d^\dagger) |0\rangle$. If we assume that the ``vacuum'' $|0\rangle$
has a positive norm, $\langle 0|0 \rangle=1$, we find
\bea
H|c\rangle &=& 0, ~~~~~~~~
H|d\rangle = |c\rangle \nonumber \\
\langle c| c\rangle &=& 0, ~~~~~~~~ \langle c| d\rangle =1, ~~~~~~ 
\langle d| d\rangle =2
\label{3djordan}\eea
We can see from eq. (\ref{3djordan}) that even if we give the vacuum a
positive norm, we are forced to have another eigenstate of the
Hamiltonian $|c\rangle$, which has zero energy and zero norm.
The other consistent way to quantize the theory is simply to take the
vacuum to have zero norm: $\langle 0|0 \rangle=0$. Since the
``vacuum'' here is really a state with a critical charge, this would
mean that the critical charge would decouple from physical
spectrum. For this reason it is natural to view the critical charge in
TMGT as being the equivalent of a singular vector in a CFT, which
would mean that the critical charge was actually
$\psi\cdot\mu=(k+2)/2$, or $\psi\cdot\mu=(k-C_2+2)/2$ in the
supersymmetric case (in the case of $SU(2)$, $C_2=2$, so in the
supersymmetric case our lowest order result would be exact, as we
found in \cite{us}). The charges with $\psi\cdot\mu =(k+1)/2$, which
are not singular vectors but do have descendants with negative norm in
the WZNW model, would seem to correspond more naturally
to super-critical charges
which have imaginary energy bound states but no state with exactly zero
energy, but this clearly cannot be correct at least for single
particle bound states. One possibility is that these charges are
super-critical for multi-particle bound states, which may form because
gauge bosons with the same charge can attract in TMGt \cite{kp}.
Also, the charges with $\psi\cdot\mu>(k+2)/2$
are also singular vectors in the WZNW model, but at levels $>1$, and
so these may correspond to critical bound states of more than one
particle in the TMGT.

The first possibility considered above, leading to
eq. (\ref{3djordan}), is what we have to  consider if we insist on adding
(super-) critical charged matter to the TMGT theory. the prototype for
this situation is
the supersymmetric
 model at level $k=C_2$, since all non-zero charges are
excluded from the unitary theory and so this is the only 
non-trivial possibility. As was pointed out in
\cite{ss}, this does not necessarily lead to a catastrophic
instability if we allow for a Hilbert space with a metric that is not
positive definite.
As we have seen in eq. (\ref{3djordan}), the
Hamiltonian becomes  non-diagonalizable, which is a familiar property of
logarithmic CFT (LCFT) \cite{gurarie}
 where the 2d Hamiltonian  is the Virasoro
 operator $L_0$ which acts on a logarithmic states as
\begin{eqnarray}
L_{0}|C> = \Delta |C>, ~~~~~~ L_{0}|D> = \Delta |D> + |C> \label{example}
\end{eqnarray}
 and norms of the states are
given by two-point correlation function, with one  zero norm state
\cite{ckt}
\begin{eqnarray}
\langle C(x) D(y)\rangle &&= 
\langle C(y) D(x) \rangle  = \frac{c}{(x-y)^{2\Delta_C }}\nonumber \\
\langle D(x) D(y)\rangle &&= 
 \frac{1}{(x-y)^{2\Delta_C}} \left(-2c\ln(x-y) + d\right)
\nonumber \\ 
\langle C(x) C(y)\rangle  &&= 0
\label{CC}
\end{eqnarray}
The resemblance between these equations and eq. (\ref{3djordan}) is
striking. Indeed, since the analysis leading to eq. (\ref{3djordan})
is not specific to the TMGT but applies to any theory 
with a critical
state, and logarithmic operators appear in CFT when the dimensions of 
two operators become degenerate, we 
can regard the Jordan block structure of LCFTs as
originating by the same mechanism, with $L_0$ as the Hamiltonian.
It is also interesting that in a WZNW model (but not in all
LCFTs)
the field
$C(x)$ which has zero norm always has to be a singular vector of
another primary field \cite{origin}, just as the 3 dimensional 
Hamiltonian can only be non-diagonalizable if there is a critical state.  
The null vectors in a WZNW model are the states 
$(E^\psi_{-1})^{k+1-2\psi\cdot\mu}|\mu\rangle =
|\mu+(k+1-2\psi\cdot\mu)\psi\rangle $, for each of which, if our
interpretation of the critical state is correct, there should be a
critical bound state of $k+1-2\psi\cdot\mu$ charged bosons and a
source with $\psi\cdot\mu'=k+1-\psi\cdot\mu$. These two states and the
logarithmic state $|D\rangle$ then satisfy
\bea
(E^\psi_{-1})^{k+1-2\psi\cdot\mu}|\mu\rangle &\sim& |C\rangle \nonumber \\
(E^{-\psi}_{1})^{k+1-2\psi\cdot\mu}|D\rangle &\sim& |\mu\rangle
\eea
So we can see that null states in the CFT on the boundary lead to
critical bound states in the TMGT in the bulk, and if we want to
include
 critical states which do not decouple from the spectrum, we end up
with a LCFT on the boundary and a similar type of theory in the bulk.
Indeed it was first noticed in \cite{cf} that in the WZNW model for
$SU(2)$ there are in general logarithmic singularities in the
conformal
blocks of primary fields with $j>k/2$. This is now known to
indicate the existence of a LCFT, and contrary to what was said in \cite{cf}
these conformal blocks can be combined to form local correlation
functions. We therefore expect that for every unitary WZNW model,
there is a logarithmic WZNW model with the same central charge. This
would be parallel to the situation 
for the minimal models, where for every ordinary CFT in the $c_{p,q}$
series there is an enlarged model with the same central charge which
is a LCFT \cite{flohr}.

\section{Conclusions}

These results are a confirmation that null vectors in CFT correspond to
critical charges in three dimensions, as was found for one particular
case in \cite{us}. We have now shown that for 
 every null vectors at level $1$
in a WZNW model 
there is a critical bound state of a single charged gluon in a TMGT,
and we expect that null vectors at higher levels will be associated
with critical multi-gluon states. It is still unclear why in the
three-dimensional picture sources with $\psi\cdot\mu=(k+1)/2$ (eg. a
source in the representation $j=(k+1)/2$ for $SU(2)$), which in the
WZNW model have
descendants with negative norm but are not equivalent to singular  
vectors of unitary representations, should also be excluded from the
spectrum. However, it is not surprising that a different mechanism
should be needed to explain the decoupling of the states which are not
singular vectors in the CFT.  
Another interesting question is the implications of these
results for non-compact groups, since bound states of charged gluons
can presumably occur in that case as well, which could lead to some
states decoupling from the spectrum, as seems to be required for the
no-ghost theorem for $SU(1,1)$ \cite{egp}.

There are a number of interesting questions that remain. It will be
interesting to check that higher-order corrections do result in the
critical charges coinciding with null vectors in the CFT. Also,
although it is very interesting that the Hamiltonians of both the TMGT
and the CFT become non-diagonalizable when critical charges are added
to the model, it is 
not yet clear 
exactly what the relation between the logarithmic operators in
CFT and the critical states in TMGT is.

{\bf Acknowledgments}

I am  grateful to Ian Kogan for very helpful discussions.

\newcommand{\NPB}[1]{ Nucl. Phys. {\bf B#1}}
\newcommand{\Ann}[1]{ Ann. Phys. {\bf #1}}
\newcommand{\CMP}[1]{ Commun. Math. Phys. {\bf #1}}
\newcommand{\PLB}[1]{ Phys. Lett. {\bf B#1}}
\newcommand{\PRL}[1]{ Phys. Rev. Lett. {\bf #1}}
\newcommand{\PTP}[1]{ Prog. Theor. Phys. {\bf #1}}
\newcommand{\MPLA}[1]{ Mod. Phys. Lett. {\bf A#1}}
\newcommand{\IJMP}[1]{ Int. J. Mod. Phys. {\bf A#1}}
\newcommand{\IJMPB}[1]{ Int. J. Mod. Phys. {\bf B#1}}
\newcommand{\CQG}[1]{ Class. Quant. Grav. {\bf #1}}
\newcommand{\PRD}[1]{ Phys. Rev. {\bf D#1}}
\newcommand{\PRB}[1]{ Phys. Rev. {\bf B#1}}
\newcommand{\JMP}[1]{ J. Math. Phys. {\bf #1}}


\begin{thebibliography}{99}
\bibitem{witten} E. Witten, \CMP{121} (1989) 351.
\bibitem{MS} G. Moore and N. Seiberg, \PLB{220} (1989) 220.
\bibitem{bpz} A.A. Belavin, A.M. Polyakov and A. B. Zamolodchikov,
 Nucl. Phys. {\bf B241} (1984), 333.
\bibitem{gko} P. Goddard, A. Kent and D. Olive, \CMP{103} (1986) 105.
\bibitem{us} I. Kogan and A.Lewis, Phys. Lett. {\bf B431} (1998), 77.
\bibitem{ss} B. Schroer and J. Swieca, \PRD{2} (1970), 2938.
\bibitem{ziegel}
 W.  Siegel, \NPB{156}, (1979), 135.
\bibitem{sch}
 J.F. Schonfeld, \NPB{185}, (1981),157.
\bibitem{tmgt}
 S. Deser, R. Jackiw and S. Templeton, \PRL{48} (1982),  975;
 Ann.Phys.(N.Y.) {\bf 140} (1982),  372. 
\bibitem{st} N. Sakai and Y. Tanii, Prog. Theor. Phys. {\bf{83}}
    (1990), 968.
\bibitem{vkpr} P. Di Vecchia, G. Knizhnik, J. L. Petersen and
P. Rossi, \NPB{253} (1985), 701.
\bibitem{acks}  G. Amelino-Camelia, I. I. Kogan, R. J. Szabo,
\NPB{480} (1996), 413.
\bibitem{gw} D. Gepner and E. Witten \NPB{278} (1986), 493. 
\bibitem{dekss} I. Kogan, JETP Lett. {\bf 49} (1989), 225;
\\M. I. Dobroliubov, D. Eliezer, I. I. Kogan,
G. W. Semenoff and R. J. Szabo, \MPLA{8} (1993), 2177.
\bibitem{fuchs} J. Fuchs, Nucl.Phys. {\bf B 286} (1987) 455.
\bibitem{kp} I. I. Kogan and I. V. Polubin, JETP Lett. {\bf 51}
(1990), 560; \PLB{252} (1990), 237.
\bibitem{gurarie} V. Gurarie, Nucl. Phys. {\bf B410} (1993), 535.
\bibitem{ckt} J.S. Caux, I.I. Kogan and A. Tsvelik,
Nucl. Phys. {\bf B} 466 (1996), 444.
\bibitem{origin} I. Kogan and A. Lewis, Nucl. Phys. {\bf B509} (1998) 687.
\bibitem{cf} P. Christe and R. Flume, Nucl. Phys. {\bf B282}  (1987) 466.
\bibitem{flohr} M. Flohr, \IJMP{12} (1997), 1943.
\bibitem{egp} J. M. Evans, M. R. Gaberdiel and M. J. Perry, hep-th/9806024

\end{thebibliography}
\end{document}